\documentclass[11pt]{article}

\usepackage[utf8]{inputenc}
\usepackage{amssymb,amsmath,amsthm}
\usepackage{ctable}
\usepackage{xspace}
\usepackage{enumerate}
\usepackage{graphicx}
\usepackage{algorithmicx}
\usepackage{algpseudocode}
\usepackage{algorithm}
\usepackage{sigsam}
\usepackage{color}
\usepackage{fancyvrb}
\DefineShortVerb[commandchars=\\\{\}]{\|}
\DefineVerbatimEnvironment{Highlighting}{Verbatim}{commandchars=\\\{\}}
\newenvironment{Shaded}{}{}
\newcommand{\KeywordTok}[1]{\textcolor[rgb]{0.00,0.44,0.13}{\textbf{{#1}}}}

\newcommand{\DecValTok}[1]{\textcolor[rgb]{0.25,0.63,0.44}{{#1}}}

\newcommand{\CharTok}[1]{\textcolor[rgb]{0.25,0.44,0.63}{{#1}}}

\newcommand{\NormalTok}[1]{{#1}}

\newtheorem{Definition}{Definition}[section]

\newtheorem{Remark}[Definition]{Remark}
\newtheorem{Example}[Definition]{Example}

\renewcommand{\FL}{}
\renewcommand{\LL}{}

\usepackage{color}

\newcommand{\flac}[2]{\lfloor\frac{#1}{#2}\rfloor}

\makeatletter
\def\maxwidth{\ifdim\Gin@nat@width>\linewidth\linewidth
\else\Gin@nat@width\fi}
\makeatother
\let\Oldincludegraphics\includegraphics
\renewcommand{\includegraphics}[1]{\Oldincludegraphics[width=0.5\textwidth]{#1}}

\renewenvironment{Shaded}{\noindent\begin{minipage}{0.6\columnwidth}\footnotesize}{\end{minipage}}

\title{Fast polynomial evaluation and composition}
\author{Guillaume Moroz}

\begin{document}
    \maketitle
    \begin{Abstract}
    The library \emph{fast\_polynomial} for Sage compiles multivariate
    polynomials for subsequent fast evaluation. Several evaluation schemes
    are handled, such as Hörner, divide and conquer and new ones can be
    added easily. Notably, a new scheme is introduced that improves the
    classical divide and conquer scheme when the number of terms is not a
    pure power of two. Natively, the library handles polynomials over gmp
    big integers, boost intervals, python numeric types. And any type that
    supports addition and multiplication can extend the library thanks to
    the template design. Finally, the code is parallelized for the divide
    and conquer schemes, and memory allocation is localized and optimized
    for the different evaluation schemes. This extended abstract presents
    the concepts behind the \emph{fast\_polynomial} library. The sage
    package can be downloaded at:
    \url{http://trac.sagemath.org/sage_trac/ticket/13358}. In Section
    \ref{scheme}, we present the notion of evaluation tree and function
    scheme that unifies and extends state of the art algorithms for
    polynomial evaluation, such as the Hörner scheme \cite{Mbook06} or
    divide and conquer algorithms \cite{Mbook06,Eacm60,BKacc75,BZsynasc11}.
    Section \ref{evaluation} reviews the different optimisations implemented
    in the library (multi-threads, template, fast exponentiation), that
    allows the library to compete with state-of-the art implementations.
    Finally, Section \ref{bench} shows experimental results.
    \end{Abstract}

    \section{Polynomial preprocessing}
    
    \label{scheme}
    
    Given a polynomial with integer, floating points, or even polynomial
    coefficients, there is several way to evaluate it. Some are better
    suited than others for specific data type. An evaluation tree specifies
    how the polynomial will be evaluated.
    
    \begin{Definition}
    An \emph{evaluation tree} $\mathcal T_p$ associated to a polynomial $p$
    is an acyclic graph with a root node $R$. Each node $N$ corresponds to a
    monomial of $p$ and has \emph{2} labels, denoted by $c(N)$, the
    \emph{coefficient} associated to $N$, and $d(N)$, the \emph{partial
    degree} associated to $N$. The result of an \emph{evaluation tree} on
    $x$ is defined recursively: \[ \mathcal T(x) = \left\{\begin{array}{ll}
        c(R)x^{d(R)}
            & \text{ if R is the only node of $\mathcal T$.}\\
        (c(R)+\sum_i \mathcal S_i(x))x^{d(R)}
            & \text{ otherwise, where $\mathcal S_i$ are the children
            tree of $R$.}\\
    \end{array}\right. \]
    \end{Definition}

    Each node of an evaluation tree is naturally associated with a term of
    the input polynomial. However, the \emph{partial degree} of a node $N$
    is not the degree of monomial associated to $N$. The degree of the
    monomial associated to $N$ is rather the sum of the partial degrees of
    its ancestors.
    
    If we order the terms of $p$ in a decreasing lexicographical ordering,
    we induce naturally an ordering on the nodes of $\mathcal T_p$. This
    ordering is also a topological ordering of $\mathcal T_p$ and will be
    denoted subsequently by $<_t$. The first node is the bigger for $<_t$
    and will have index $0$. The last node is the root of the tree and will
    have index $n$. In particular, all the children of a node of index $i$
    have an index lower than $i$.
    
    \subsection{Function scheme}
    
    A way to define an evaluation scheme for univariate polynomials is to
    use a \emph{function scheme}.
    
    \begin{Definition}
    Let $f:\mathbb N \rightarrow \mathbb N$ be a function such that
    $0<f(k)\leq k$ for all $k\geq 1$. Let $p$ be a univariate polynomial of
    degree $n$. We define recursively the evaluation tree $\mathcal T_p^f$
    associated to the \emph{function scheme} $f$.
    
    If $p$ has one term, then $\mathcal T_p^f$ is reduced to one node of
    coefficient and degree those of the term in $p$. Otherwise, $p$ can be
    written uniquely $p(x) = a(x)x^{f(n)} + b(x)$. The evaluation tree
    $\mathcal T_p^f$ is obtained by adding the tree $\mathcal T_a^f$ as a
    child of the root of the tree $\mathcal T_b^f$.
    \end{Definition}

    Most classical schemes such as Hörner \cite{Mbook06} or Estrin (divide
    and conquer \cite{Mbook06,Eacm60,BKacc75,BZsynasc11}) schemes can be
    described with simple function schemes:
    
    \begin{align*}
    \text{Direct: } & D(k) = k & \text{Hörner: } & H(k) = 1 
    & \text{Estrin: } & E(k) = 2^{\lfloor\log k\rfloor}\\
    \end{align*}
    
    \begin{Example}
    Let $p$ be the polynomial $3x^8-x^7+2x^6+x^5-4x^4+9x^3-3x^2-2x+1$. Then
    the following trees are all evaluation trees of $p$, with different
    evaluation scheme.
    \end{Example}

    \ctable[pos = H, center, botcap]{cc}
    {
    }
    {
    \FL
    \includegraphics{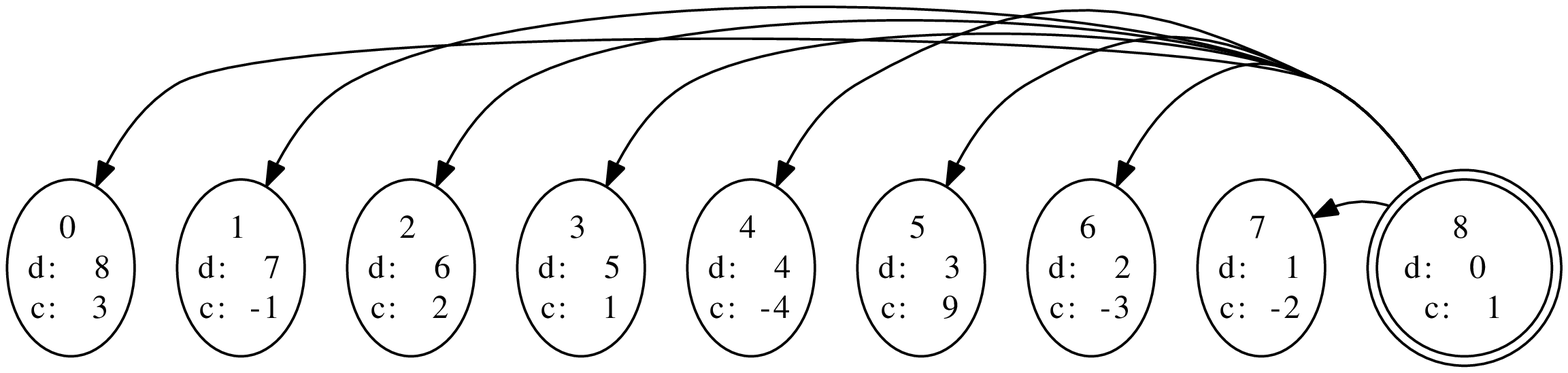} & \includegraphics{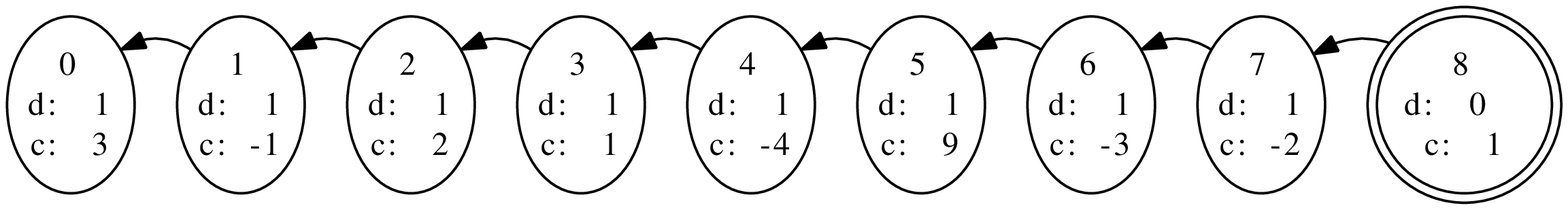}
    \\\noalign{\medskip}
    Direct scheme & Hörner scheme
    \LL
    }
    
    \ctable[pos = H, center, botcap]{c}
    {
    }
    {
    \FL
    \includegraphics{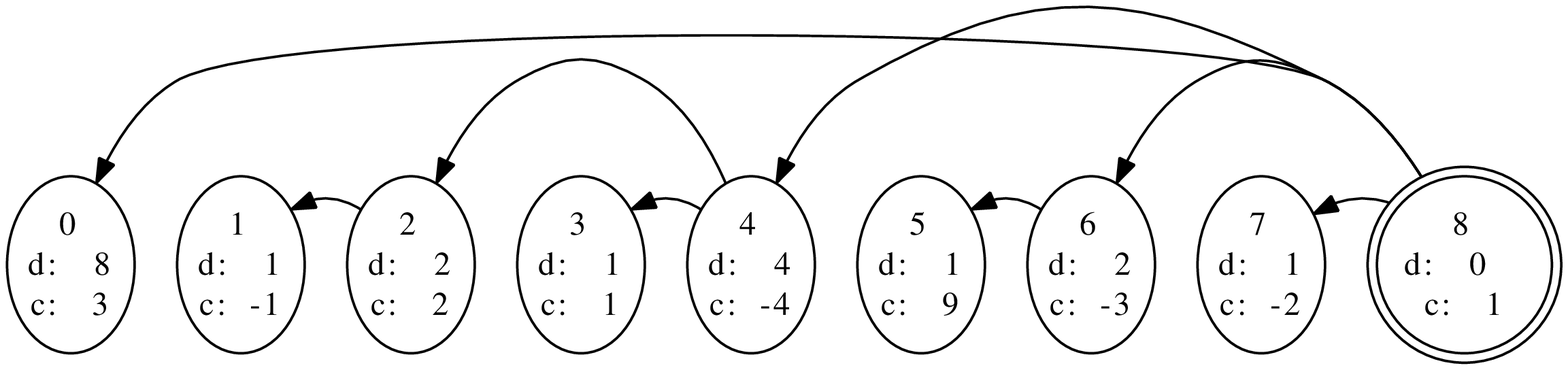}
    \\\noalign{\medskip}
    Estrin scheme
    \LL
    }
    
    \begin{Remark}
    For multivariate polynomials, the function scheme can be applied
    recursively to each variable.
    \end{Remark}

    \begin{Remark}
    Function schemes can be defined and used in \emph{fast\_polynomial}
    library, as documented in the module \emph{method}. It is thus possible
    to combine easily different schemes. For example, let $f$ be the
    function $f(k)=2^{\lfloor\log k\rfloor}$ if $k>10$ and $f(k)=1$
    otherwise. The corresponding evaluation tree is a divide and conquer
    scheme for the upper part and a Hörner scheme for the sub polynomials of
    degree less than $10$.
    \end{Remark}

    \subsection{A new balanced divide and conquer scheme}
    
    The Estrin scheme is a divide and conquer algorithm well suited to
    evaluate polynomials on elements whose size increases linearly with each
    multiplications (\cite{BKacc75,BZsynasc11}). These elements include
    multiple precision integers or univariate polynomials. However, the
    computation time of evaluating $\mathcal T_p^E$ reaches thresholds when
    the number of terms of $p$ is a pure power of $2$ (see Figure
    \ref{integer} in Section \ref{bench}).
    
    We introduce in this library a new evaluation scheme that avoids the
    time penalty of the classical divide and conquer. It is defined by the
    \emph{balanced} function scheme.
    
    \[ \text{Balanced: } B(k) = \flac{k}{2}\]
    
    \begin{Example}
    {[}continued{]} The balanced divide and conquer evaluation trees
    contains lower partial degrees in this example.
    \end{Example}

    \ctable[pos = H, center, botcap]{c}
    {
    }
    {
    \FL
    \includegraphics{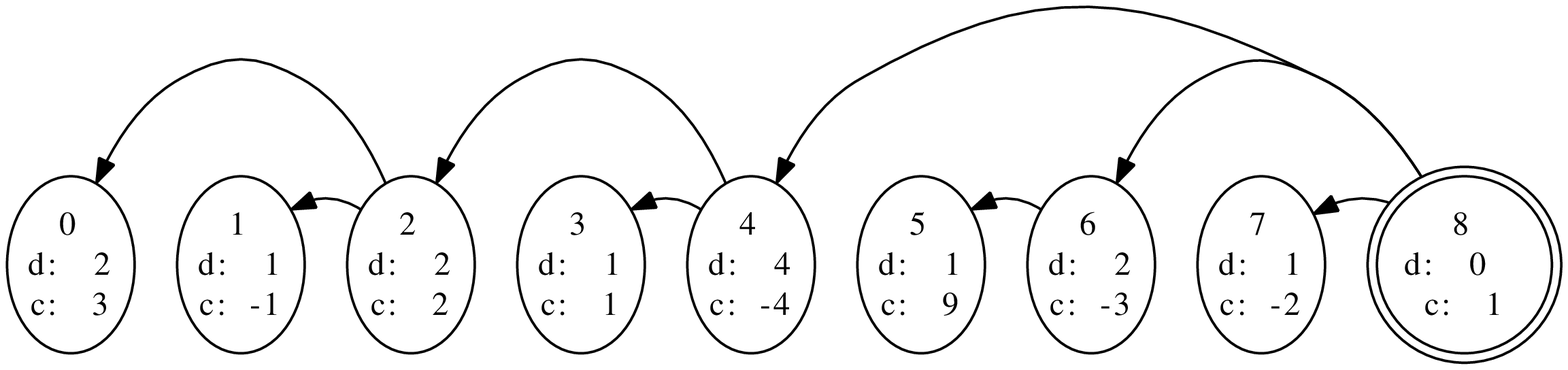}
    \\\noalign{\medskip}
    Balanced divide and conquer scheme
    \LL
    }
    
    \subsection{Lazy height}
    
    We associate to each node of the tree a \emph{lazy height}, that will
    determine the number of temporary variables required during the
    evaluation. In particular, the \emph{lazy height} must be kept as low as
    possible. Classically, the height of a node is always greater then the
    height of its children. In our case, the \emph{lazy height} of a node is
    greater than the \emph{lazy height} of its children only if it has two
    or more children. In particular, this ensures us that for any tree, the
    maximal \emph{lazy height} is at most logarithmic in the number of
    nodes.
    
    \begin{Definition}
    Let $N$ be a tree node. The \emph{lazy height} of $N$, denoted $lh(N)$,
    is defined recursively. Let $C_1,...,C_k$ be the child nodes of $N$ such
    that $c_1>_t\cdots>_tc_k$. \[ lh(n) = \left\{ \begin{array}{cl}
                0        &\text{ if $N$ has $0$ or $1$ child.}\\
                \displaystyle\max_{2\leq i\leq k}(lh(C_i))+1 &\text{ otherwise.}
              \end{array}\right. \]
    \end{Definition}

    \begin{Example}
    Consider again the polynomial
    $p = 3x^8-x^7+2x^6+x^5-4x^4+9x^3-3x^2-2x+1$. In the case of Hörner
    scheme, the maximal \emph{lazy height} of the associated evaluation tree
    is $0$, whereas its classical height is $8$. The \emph{lazy height}
    associated to the Direct scheme is $1$. And we can check that the Estrin
    scheme and the Balanced scheme have both maximal \emph{lazy heights}
    $1$.
    \end{Example}

    \section{Evaluation}
    
    \label{evaluation}
    
    \subsection{Coefficients walk}
    
    Once the tree data structure has been computed, the evaluation can be
    done efficiently. If $p$ is a univariate polynomial of degree $n$, we
    can use the following pseudo-code.
    
    \begin{Shaded}
    \begin{Highlighting}[]
    \KeywordTok{for} \NormalTok{i }\CharTok{from} \DecValTok{0} \NormalTok{<= i < n:}
        \NormalTok{N = nodes[i]}
        \NormalTok{c, d, h = N.coefficient, N.partial_degree, N.lheight}
        \NormalTok{p = (m[h] + c)*x^d}
        \NormalTok{m[h] = }\DecValTok{0}
        \KeywordTok{if}   \NormalTok{i == n: }\KeywordTok{return} \NormalTok{p}
        \KeywordTok{elif} \NormalTok{i <  n: m[N.parent.lheight] += p}
    \end{Highlighting}
    \end{Shaded}
    \begin{minipage}{0.4\columnwidth}
    If the values $x^d$ have been precomputed (see next Section), each step
    costs one multiplication and one addition. The mutable variables are $p$
    and $m[0],...,m[L]$, where $L$ is the lazy height of the root node. Their
    number is at most $O(\log n)$.
    \end{minipage}
    
    \subsection{Powers computation}
    
    The powers $x^d$ appearing in the evaluation loop can be computed
    several times for the same $d$. In order to optimize the evaluation,
    these powers can be precomputed using fast exponentiation methods.
    
    \newcommand{\Dd}{$1,\ldots,n$}
    
    \newcommand{\Dh}{$1$}
    
    \newcommand{\De}{$2^k$}
    
    \newcommand{\Db}{$\flac n{2^k},\flac n{2^k}+1$}
    
    \newcommand{\logn}{$0\leq k \leq \log n$}
    
    Assume that $p$ is a dense univariate polynomial of degree $n$. Table
    \ref{powers} shows that the balanced scheme, as well as the Estrin
    scheme, require at most a logarithmic number of different powers to
    compute.
    
    \scriptsize
    
    \ctable[caption = {Degrees appearing in the evaluation tree of a dense
    polynomial of degree $n$. \label{powers}},
    pos = H, center, botcap]{cccc}
    {
    }
    {
    \FL
    \parbox[b]{0.15\columnwidth}{\centering
    Direct
    } & \parbox[b]{0.15\columnwidth}{\centering
    Hörner
    } & \parbox[b]{0.15\columnwidth}{\centering
    Estrin
    } & \parbox[b]{0.15\columnwidth}{\centering
    Balanced
    }
    \ML
    \parbox[t]{0.15\columnwidth}{\centering
    \Dd
    } & \parbox[t]{0.15\columnwidth}{\centering
    \Dh
    } & \parbox[t]{0.15\columnwidth}{\centering
    \De
    } & \parbox[t]{0.15\columnwidth}{\centering
    \Db
    }
    \\\noalign{\medskip}
    \parbox[t]{0.15\columnwidth}{\centering
    } & \parbox[t]{0.15\columnwidth}{\centering
    } & \parbox[t]{0.15\columnwidth}{\centering
    \logn
    } & \parbox[t]{0.15\columnwidth}{\centering
    \logn
    }
    \LL
    }
    
    \normalsize
    
    \subsection{Template system and multi-thread}
    
    The code is written with templates, and is specialized for different
    \emph{C/C++} object. This allows the library to compete with
    state-of-the art ad hoc implementations, and to be easily extended with
    new numeric types (see \emph{interfaces/README} in the package).
    
    Moreover, the evaluation tree can be evaluated with multiple threads in
    parallel. The parallelization mechanism is implemented with openMP
    directives.
    
    \section{Benchmarks}
    
    \label{bench}
    
\begin{figure}[htbp]
    \centering
    \includegraphics{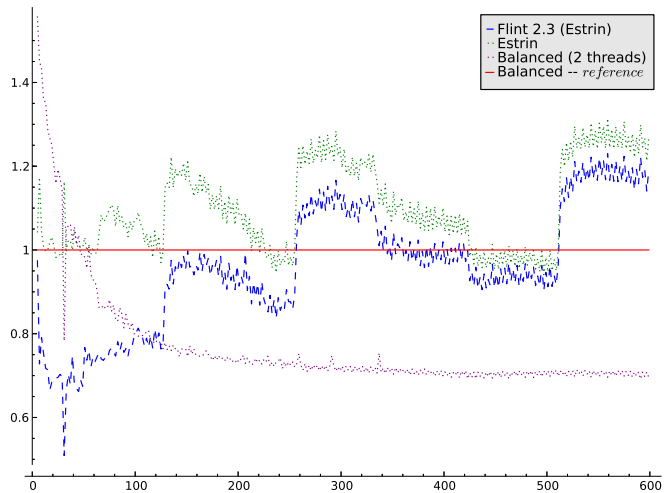}
    \caption{\scriptsize Comparison of the Balanced scheme with the Estrin
    scheme, the Balanced scheme with 2 threads, and the state of the art
    Flint library. The abscisse represents the degree of the polynomial $p$,
    the bitsize of its coefficients, and the bitsize of the integer on which
    it is evaluated. The ordinate represents the computation time for the
    different methods divided by the computation time for the Balanced
    scheme. \label{integer}}
    \end{figure}

    The Figure \ref{integer} shows the performance of the balanced scheme
    implemented in \emph{fast\_polynomial} for the evaluation over multi
    precision integers. We see in particular that the balanced scheme
    doesn't suffer the staircase effect shown by the classical divide and
    conquer algorithms for pure powers of $2$. The results suggest also that
    an implementation of the balanced scheme directly in Flint could improve
    the polynomial composition and evaluation over big integers in some
    cases.
    
    \footnotesize
    \bibliographystyle{alpha}
    \bibliography{biblio}

\begin{thebibliography}{Mul06}

\bibitem[BK75]{BKacc75}
Richard~P Brent and HT~Kung.
\newblock 0 ((n log n) 3/2) algorithms for composition and reversion of power
  series.
\newblock {\em Analytic computational complexity}, pages 217--225, 1975.

\bibitem[BZ11]{BZsynasc11}
Marco Bodrato and Alberto Zanoni.
\newblock Long integers and polynomial evaluation with estrin's scheme.
\newblock In {\em Symbolic and Numeric Algorithms for Scientific Computing
  (SYNASC), 2011 13th International Symposium on}, pages 39--46, 2011.

\bibitem[Est60]{Eacm60}
Gerald Estrin.
\newblock Organization of computer systems: the fixed plus variable structure
  computer.
\newblock In {\em Papers presented at the May 3-5, 1960, western joint
  IRE-AIEE-ACM computer conference}, pages 33--40. ACM, 1960.

\bibitem[Mul06]{Mbook06}
Jean-Michel Muller.
\newblock {\em Elementary functions}.
\newblock Computer Science. Birkh{\"a}user Boston, 2006.

\end{thebibliography}
\end{document}